\documentclass[aps,pra,twocolumn,showpacs,floatfix]{revtex4-1}
\usepackage{graphicx}
\usepackage{hyperref}
\begin{document}
\title{Effect of dissipative environment on
collapses and revivals of a non-linear quantum
oscillator}
\author{Maninder Kaur}
\email{ mannu\_711@yahoo.co.in}
\affiliation{Department of Physics, Guru Nanak Dev
University, Amritsar, Punjab-143005, India}
\author{Bindiya Arora}
\email{bindiya.phy@gndu.ac.in}
\affiliation{Department of Physics, Guru Nanak Dev
University, Amritsar, Punjab-143005, India}
\author{ Arvind}
\email{arvind@iisermohali.ac.in}
\affiliation{Department of Physical Sciences,
Indian Institute of Science Education
and Research (IISER) Mohali, PO Manauli Punjab
140306, India}
\begin{abstract}
We study the dissipative dynamics of a wave packet passing
through two different non-linear media.  The effect of
dissipation on the  phenomenon of collapses and revivals of
a wave packet as it evolves in a Kerr-type non-linear medium
(represented by the Hamiltonian $({a}^{\dag} a)^2$) is
investigated.  We find that partial revivals do take place
when dissipation values are moderate.  For a certain regime
of parameters we find a solution where revivals do not die
even in the presence of dissipation and the non-linearity
appears to compensate for the energy and coherence loss.  We
consider the next order non-linearity, represented by the
Hamiltonian $({a}^{\dag} a)^3$, where we observe the
phenomena of super revivals.  The effect of dissipation in
this case has an additional feature of number dependence for
the displaced number states.  While our simulations explore
the degree to which  the phenomena of collapses and revivals
degrades in a dissipative environment, we also discovered
the presence of a situation where degradation is minimal.
\end{abstract}
\maketitle
\section{Introduction}
\label{sec1}
Dynamics of quantum systems in complex
environments plays an important role in diverse
fields of science~\cite{sw1,sw2,sw4,sw5,sw7,sw3}.
A laser beam that is quantum mechanically
represented by a coherent state,  while passing
through non-linear media, can undergo a variety of
complex transformations  including collapses and
revivals of the quantum state. Over and above any
systematic evolution (linear or non-linear),
dissipation is always present, and therefore a
realistic study must include the effect of a
dissipative environment~\cite{sw3,bookwm,pfoldi}.
Classically the dissipative effects lead to
diminishing of the amplitude, however, since the
interactions occur at atomic scales, quantum
effects are important.  An important set of
quantum states is  obtained by a phase space
displacement of the number states (states with a
fixed number of quanta
$n$)~\cite{cwp1,cwp3,cwp4}.  A reasonable
model for dissipation on the other hand is to
model the dissipative medium as a set of
oscillators that interact with the
system~\cite{bookwm}.  For linear as
well non-linear media the initial motion of the
wave packet is periodic with the period of motion
termed as the `classical period' $T_{\rm
cl}$~\cite{Per93}, corresponding to the natural
frequency of the underlying harmonic oscillator.
For a linear medium the periodic character of the
motion is preserved; however, for a  non-linear
medium,  after a few cycles quantum interference
takes over leading to a significant spread of the
wave packet.  The wave packet is no longer
recognizable as a packet and is said to have
``collapsed''. As time advances, it resurrects
itself leading to its ``revival''. The time at
which this revival takes place is called the
`revival time' $T_{\rm rev}$~\cite{ KS96,
AR95, Ave92}. At times that are rational fractions
of $T_{\rm rev}$, the wave packet turns into a
series of subsidiary packets and the phenomenon is
called fractional revival.  For the case of third
order non-linearity, for times beyond revival
times, a new sequence of full and fractional
revivals start, which are characterized by a
longer time scale called the `super revival time'
$T_{\rm sr}$.  The phenomenon of collapses and
revivals of the wave packets was first discussed
in~\cite{s&s} and thereafter, has been studied by
a number of
authors~\cite{2jm,3jm,4jm,5jm,sud1,12jm, 14jm,
15pra}.  In most of these  studies the  medium is
assumed to be non-dissipative.  However, for
situations of practical interest the medium can
not be assumed to be non-dissipative or ideal.
How to incorporate the effects of dissipation on
the phenomena of collapses and revivals of wave
packets is the central question that we address in
the present paper.

Unlike in classical mechanics, the dissipative terms cannot
be directly incorporated in quantum equations of motion as
this leads to the decay of the Heisenberg uncertainty
relation which is absurd. To overcome this difficulty
various approaches have been proposed including those
involving complex Hamiltonians~\cite{oa1,oa2,oa3,new_ref6}.
We consider the model for dissipation where the system
interacts with the environment via an interaction
Hamiltonian. The composite system consisting of the system
of interest plus the environment evolves unitarily and the
environment degrees of freedom are traced over to arrive at
the dynamics of the system alone~\cite{sw3,bookwm}.
Concrete physical models involving nonclassical light and
dissipative media in cavity QED, quantum wells  and plasma
systems have been
proposed~\cite{new_ref1,new_ref2,new_ref3}.  Rydberg atom in
a micocavity weakly excited in the strong coupling limit has
been investigated from this point of view~\cite{new_ref5}.
Quantum correlations reflected in quantum discord for open
system has also been studied~ \cite{new_ref4}.

Our work embarks upon a study of the dynamics of a
quantum wave packet through a non-linear medium in
the presence of dissipation.  In particular we aim
to study the effects of dissipation on the
phenomenon of wave packet revival and
super-revival.  We consider a single-mode radiation
field as our system and  take coherent states
and displaced number states as our wave packets.
Such packets can arise  in many experimental
situations~\cite{dns1,cwp,cwp11}.  For non-linear
media we first consider a Kerr medium represented by the
Hamiltonian $({a}^{\dag} a)^2$. The dynamics of the coherent
wave packet in this case gives rise to revivals and
collapses of the wave packet.  To include the effects of
dissipation we write down a master equation corresponding to
a situation where the single mode radiation field interacts
with a set of harmonic oscillators representing a thermal
bath.  Specifically we have assumed that we have
single mode radiation in the visible region (e.g.
$\lambda=600$ nm red light). From the master equation, we
numerically calculate the time evolution of the expectation
value of the amplitude of the radiation field.  Next, we
study the one order higher non-linearity represented by a
term proportional to $({a}^{\dag} a)^3$  in the Hamiltonian.
This medium is  a favorable system to observe the super
revivals of the wave packet and we study these revivals
under dissipation. Moreover, the dynamics of coherent wave
packet in this medium provides the opportunity to analyze
the effect of dissipation caused by the medium on the
displaced number states.  An important result we obtain is
that the revivals as well as super-revivals can tolerate a
certain amount of dissipation and thus can be observed in
experiments where dissipation is inevitable. The tolerance
to dissipation, however, depends upon the type and strength
of the non-linearity.

A noteworthy result is that for a
certain parameter regime for the Kerr-type
non-linearity,  revivals survive the effects of
dissipation over a very long period of time. The
simulations indicate that the loss of energy is
compensated by the non-linearity. This is a
situation which resembles that of solitons in
classical systems. Furthermore, along with energy
loss being compensated by non-linearity, the
coherence is also preserved and the revivals are
almost complete. To the best of our knowledge this
is the first observation of this phenomenon.  In
the classical domain solitons are wave forms which
travel in a lossy non-linear medium without losing
shape and amplitude. Solitons have
been extensively discussed for optical
situations~\cite{sol_prep} and also for Kerr-type
non-linearity~\cite{sol_kerr1,sol_kerr2}. 
Solitons in physical system play an important
role~\cite{soliton_phys} Interaction of solitons with atomic
systems has also been studied~\cite{new_ref_sol}. 

Strictly speaking solitons are
associated with situations where dispersion is
compensated by nonlinearity, however, the control
of dissipation is also sometime called soliton
and that is why we call our situation soliton
like~\cite{ws}.

There have also been quantum generalizations of  solitons
for optical situations~\cite{sol_nature}. However in most of
these cases the starting point is classical solitons which
are then quantized. In our case the situation is very
different, where we are  working with an open quantum system
with non-linearity being present and the ``soliton-like''
solutions automatically emerge for certain parameter
regimes. These results are valid within the approximations
and the assumptions of the dissipation model that we have
used and are related to solitons only in the restricted
sense as used in~\cite{ws}.

The paper is organized as follows:
Section~\ref{sec2} provides the framework for our
study with the model of
dissipation discussed in Section~\ref{sys_bath},
the kind of wave packets
studied in Section~\ref{packets}  and  various
time scales pertaining to
the time evolution of wave packets in non-linear
media taken up in Section~\ref{time_scales}.
In Section~\ref{sec3}, we present the
results for the time evolution of the expectation
value of the amplitude operator in a dissipative
environment. Specifically, Section~\ref{sec52}
contains results for the effect of dissipation for the
case  Kerr non-linearity and the ``soliton-like''
situation is discussed in~\ref{sec33}, while
Section~\ref{sec53} contains results for the
higher order non-linearity.  In section~\ref{concl}
we present some concluding remarks.
\section{Dissipation Model and Revival times for the
non-linear oscillator}
\label{sec2}
\subsection{System bath model}
\label{sys_bath}
We
study system-environment interaction in the
Born-Markov  approximation which yields the
quantum master equation for the system.
Consider a quantum system $S$ interacting with its
environment $E$, which can be treated as a heat
bath or reservoir. The total (system +
environment) Hamiltonian for this system can be
written in the form~\cite{sw3}
\begin {equation}
H=H_{S}+H_{E}+H_{\rm int}~\label{H}
\end {equation}
where $H_S, H_E,$ and $H_{\rm int}$
describe the Hamiltonian  for the system,
environment and interaction between them,
respectively. For one mode radiation field
represented by bosonic annihilation
operator $a$ with
$[a,a^{\dag}]=1$,
traversing through a non-linear medium the above
Hamiltonians are given by the following
expressions:
\begin{eqnarray}
H_{S}&=&\hbar\omega_{0}{a}^{\dag} a + H_{\rm NL}\\
H_{E}&=&
\sum_{j}\hbar\omega_{j}{e}_{j}^{\dag}
e_{j}\nonumber\\
 H_{\rm int}&=& \sum_{j}
\hbar k_{j}[a{e}_{j}^{\dag}+{a}^{\dag} e_{j}].
\label{hamil}
\end{eqnarray}
Here $\omega_{0}$ is the frequency corresponding
to the harmonic oscillator representing the mode
and $H_{\rm NL}$ is the non-linear part of  the
Hamiltonian  being considered.  The dissipation is
modeled as a set of harmonic oscillators with
frequencies $\omega_{j}$, while  ${e}_{j}^{\dag}$ and
$e_{j}$ are the creation and annihilation
operators of the environment.  The interaction
Hamiltonian is in the rotating wave
approximation~\cite{sw3}  and $k_{j}$
represents the coupling strength of the $j$th bath
mode with the system.

In the Schr\"odinger picture, reduced density matrix
$\rho_{S}$ for the system is described by the
master equation~\cite{sw3,bookwm,pfoldi18,sw4,
pfoldi20,pfoldi21}
\begin{eqnarray}
\frac{d\rho_{S}}{dt}&=&-\frac{\iota}{\hbar}[H_{S},\rho_{S}]
+ \gamma (N+1)[a \rho
{a}^{\dag}-\frac{1}{2}({a}^{\dag}a\rho+\rho
{a}^{\dag} a)]\nonumber\\
&&
+\gamma N [{a}^{\dag}
\rho a-\frac{1}{2}(a {a}^{\dag} \rho +\rho a
{a}^{\dag})].
\label{fme}
\end{eqnarray}
Here the parameter
$N=\left[\exp\left(\frac{\hbar\omega_{0}}{k_{B}T}\right)-1\right]^{-1}$
represents the number of bath quanta at frequency
$\omega_{0}$ for a bath in thermal equilibrium at
temperature $T$ and  $\gamma$ is the damping rate.  We refer
the reader to reference ~\cite{bookcarmi} for the detailed
derivation.  For cases where $\hbar\omega_{0}>>k_{B}T$ (for
instance, a photon of red light traversing an environment at
room temperature), the energy will only flow from the system
to the environment. This leads to further simplification of
the master equation where we drop the last term in
Equation~(\ref{fme})~\cite{pfoldi} to obtain
\begin{equation}
\frac{d\rho_{S}}{dt}=-\frac{\iota}{\hbar}[H_{S},\rho_{S}]
+ \gamma [a \rho
{a}^{\dag}-\frac{1}{2}({a}^{\dag}a\rho+\rho
{a}^{\dag} a)].\label{dme}
\end{equation}
The system Hamiltonian $H_{S}$ has two parts, the first
being the linear part represented by
$H_{S_0}=\hbar\omega_0 a^{\dag} a$ and the
second being the non-linear part
$H_{\rm NL}$ leading to
\begin{eqnarray}
\frac{d\rho_{S}}{dt}=-\frac{\iota}{\hbar} [H_{S_0},\rho_{S}]
-\frac{\iota}{\hbar} [H_{\rm NL},\rho_{S}]+\nonumber\\
\gamma [a \rho
{a}^{\dag}-\frac{1}{2}({a}^{\dag}a\rho+\rho
{a}^{\dag} a)].\label{sdme} \end{eqnarray}
We consider two cases of non-linear terms
$H_{\rm NL}$ as given below
\begin{eqnarray}
H_{\rm NL1}&=& \hbar b_{1} ({a}^{\dag} a)^{2},\nonumber \\
H_{\rm NL2}&=&\hbar b_{2} ({a}^{\dag} a)^{3}\label{b1-b2},
\end{eqnarray}
where $b_{1}$ and $b_{2}$
define the strength of nonlinearity and are in units of frequency.

Equation~(\ref{sdme}) can be written in  the
super-operator form as
\begin{equation}
\frac{d\rho_{S}}{dt}=\mathcal{L}\rho_{S},~\label{lme}
\end{equation} where $\mathcal{L}$ is the
Liouvillian  given by

\begin{equation}
\mathcal{L}=-\frac{\iota}{\hbar} [H_{S_0}
+H_{\rm NL},\rho_{S}]
\nonumber
+\gamma[a
\rho a^{\dag}-\frac{1}{2} (a^{\dag}
a\rho -\rho a^{\dag}a)]~\label{co}
\end{equation}
For the case of a time independent Hamiltonian the
solution of Equation~(\ref{lme})  can be formally
written as
\begin{equation}
\rho_{S}(t)=\exp[\mathcal{L}(t)]\rho_{S}(t_{0}).~\label{slme}
\end{equation}
From the above  equation one can compute the time
evolution of the expectation value of any physical
quantity of interest.  In particular, when we have
a  coherent wave packet, we would like to
calculate the time varying expectation value of
the wave packet amplitude $\langle a(t) \rangle$.
\subsection{The wave packets}
\label{packets}
Coherent wave packets that we consider as initial states
are obtained
by phase space displacements of  the vacuum
state of the harmonic oscillator~\cite{cwp1, cwp2, cwp3,
cwp4}, 
\begin{eqnarray}
\vert\alpha\rangle &=& D(\alpha)\vert0\rangle\nonumber \\
D(\alpha)&=& \exp(\alpha{a}^{\dag}- \alpha ^ {*} a)
\label{disp}
\end{eqnarray}
These are well known coherent states parameterized by the
complex parameter $\alpha$ and typically represent laser
light in quantum optics.  For the corresponding harmonic
oscillator, wave function for a coherent state is a Gaussian
and the time evolution corresponds to a Gaussian wave packet
oscillating in space with amplitude governed by $\vert
\alpha \vert$.

The action of the displacement operator
$D(\alpha)$ on number states $|n\rangle$(states
with fixed number of photons) gives
rise to displaced number states given by
\begin{equation}
\vert\alpha, n \rangle=D(\alpha)\vert n\rangle
\label{dns}
\end{equation}
For the case of higher order non-linearity, 
we  will also consider these states as initial states, in
addition to coherent states.
Starting with the initial density operator
corresponding to the system state we will
numerically compute the solution of
Equation~(\ref{slme}) and then compute the relevant
expectation values.
\subsection{Revival time scales}
\label{time_scales}
For non-linear oscillators when
the Hamiltonian commutes with the number operator,
the eigen vectors are same as those of the  underlying
linear oscillator.
The energy of
the $n^{\rm th}$ level on the other hand
can be  expanded as  a
Taylor series~\cite{5jm} around the central energy
$n_0$ when the energy spread is not too much as
follows
\begin{eqnarray}
E_n\approx E_{n_{0}} &+& (n-n_0)E_{n_0}'
+\frac{1}{2}(n-n_0)^2E_{n_0}''\nonumber \\
&&
+\frac{1}{6}(n-n_0)^3E_{n_0}'''+.....
\end{eqnarray}
where the primes over the energy
terms represent derivatives with respect to $n$ which
define the various time scales (although $n$ is discrete, it
is useful to treat it as a continuous variable in this
analysis). The first time
scale is given by
\begin{equation}
T_{\rm cl} =
\frac{2\pi\hbar}{E'_{n_0}} \end{equation}
and is the `classical' time period for the
shortest closed orbit~\cite{Per93}. It controls the
initial behavior of the wave packet.
The second time scale
\begin{equation}
T_{\rm rev}=\frac{2\pi\hbar}{\frac{1}{2}E''_{n_0}}
\end{equation} is the revival  time~\cite{KS96,
AR95}. It governs the appearance of the fractional
revivals and the full revivals. The third time
scale
 \begin{equation} T_{\rm sr} =
\frac{2\pi\hbar}{\frac{1}{6}E'''_{n_0}} \end{equation}
is the super revival time. It is a larger time
scale as compared to the classical time period and
the revival time. 
These time scales become relevant
depending upon the kind of non-linearity under
consideration.

For the Kerr-type non-linearity
the Hamiltonian is given by
\begin{equation}
H_{S_{1}}=\hbar\omega_{0}{a}^{\dag} a +\hbar b_{1}
({a}^{\dag} a)^{2}
\end{equation}
This Hamiltonian has the same eigen states
as the original (linear) oscillator
namely,
$\vert n\rangle$, and the corresponding energy
eigen values are given by
\begin{equation}
E_{n_{1}}=\hbar\omega_{0}n_{1} + \hbar
b_{1} n_{1}^{2}
\end{equation}
It is clear that the revival time is an important
parameter here
and is given by
\begin{equation}
T_{\rm rev}=\frac{2\pi\hbar}{\frac{1}{2}(2b_{1} \hbar)}=\frac{2\pi}{b_{1}}.\label{trvn}
\end{equation}
Similarly, for a non-linear medium
with non-linearity proportional to $({a}^{\dag}
a)^{3}$ the Hamiltonian and energy are given as
\begin{equation}
H_{S_{2}}=\hbar\omega_{0}{a}^{\dag} a + \hbar b_{2}
({a}^{\dag} a)^{3}
\end{equation} and corresponding energy eigen
states are given by
\begin{equation} E_{n_{2}}=\hbar\omega_{0}n_{2} + \hbar
b_{2} n_{2}^{3}.\nonumber \end{equation} We can
compute the revival time as
\begin{equation}
T_{\rm rev}=\frac{2\pi\hbar}{\frac{1}{2}(6b_{2}
n_{2} \hbar)}=\frac{2\pi}{3b_{2}n_{2}}~\label{trev}
\end{equation} and the super-revival time as
\begin{equation} T_{\rm sr} =
\frac{2\pi\hbar}{\frac{1}{6}(6 b_{2} \hbar)}=\frac{2\pi}{b_{2}}.~\label{tsr}
\end{equation}
It is evident that  the super-revival time becomes
relevant in this case because of the presence of
higher order non-linear terms in the Hamiltonian.
\begin{figure}[h!]
\centering
\includegraphics[scale=1]{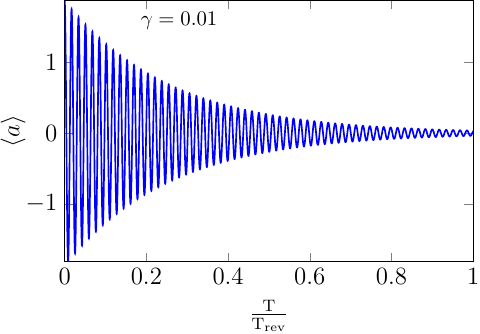}
\caption{The real part of  expectation value $\langle a\rangle$
plotted as a function of time for coherent wave
packet of red light in the presence of dissipation ($\gamma=0.01$ a.u.) and without any non-linearity. 
Time axis is normalized with respect to revival time 
$T_{\rm rev}$.}
\label{fig-1}
\end{figure}
\section{Results and discussion}
\label{sec3}
\begin{figure}[t]
\centering
\includegraphics[scale=1]{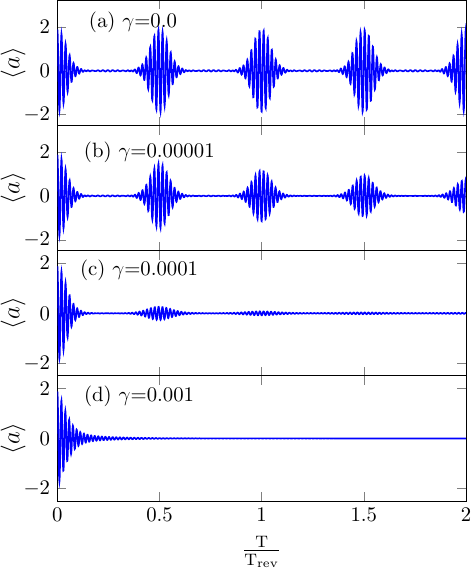}
\caption{Collapse and revival of a coherent wave
packet of red photon passing through medium with
Kerr type non-linearity i.e. proportional to
$(a^{\dag}a)^2$ for different values of
$\gamma$. (a) $\gamma=0.0$ a.u., (b) $\gamma =0.00001$ a.u.,
(c) $\gamma=0.0001$ a.u.
and (d) $\gamma= 0.001$ a.u.}
\label{fig-2}
\end{figure}
\begin{figure}[h!]
\centering
\includegraphics[scale=1]{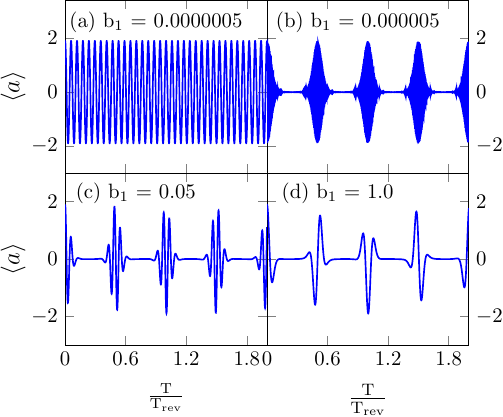}
\caption{Effect of weak and strong non-linearity
represented by parameter $b_1$, on the revival behavior
for a coherent wave packet of red photon passing
through medium without dissipation. The values of
non-linearity are (a) $b_1=0.0000005$ a.u., (b)
$b_1=0.000005$ a.u., (c)
$b_1=0.05$ a.u. and (d)$ b_1= 1.0$ a.u. }
\label{fig-3}
\end{figure}
We carry out the numerical solution of Equation~(\ref{slme})
to compute the time evolved state of the system. We have
evaluated all the physical quantities in atomic units (a.u.)
by substituting $\hbar$ and $k_{B}$ equal to $1$.  These
are convenient units and their use is a standard practice in
atomic physics community.  The simulations are done for a
variety of parameters to explore the effects of dissipation
on the process of collapses and revivals for both types of
non-linearities.  We finally calculate the expectation value
of oscillator amplitude $\langle a\rangle$ as a function of
time. Throughout this paper when  plot $\langle
a\rangle$ in different circumstances, we mean its real part.

The time step for evolution in the simulations is chosen to
be much smaller than the classical time period which in turn
is smaller than the revival time and the super revival time.
The parameters in the numerical calculations are chosen to
match a physically viable situation corresponding to red
light with angular frequency $\omega_{0}=0.15 \pi$ a.u.
($\lambda=600$ nm), passing through a non-linear medium. The
initial amplitude of the coherent state pertaining to red
light is taken to be $\alpha=1.9$.  The damping parameter
$\gamma$ represents the overall coupling strength between
the system and environment and $\gamma=0.00001$ a.u., $
0.0001$ a.u. and $0.001$ a.u.  correspond to the weak,
intermediate and strong damping respectively ($\gamma$ value
of $0.0001$ a.u. corresponds to $6.58\times 10^{11}$Hz).
The strength of the non-linearity in the medium is defined
by parameters $b_1$ and $b_2$ appearing in
Equation~(\ref{b1-b2}).  We study the effect of
system-environment coupling on collapses and revivals of the
wave packet by varying the dissipation parameter $\gamma$.
In addition to this, we have  also studied the behavior of
revival and collapses as a function of non-linearity of the
medium.

In the rest of this section, we first present the
results obtained for dynamics of a coherent wave
packet traversing a dissipative medium in the
absence of any non-linearity.  Thereafter, we
discuss the findings for the media with
non-linearity proportional to $({a}^{\dag}a)^{2}$
and $({a}^{\dag}a)^{3}$.  In case of the non-linear
term proportional to $(a^{\dag}a)^2$, the medium
behaves as an optical Kerr medium. The initial
coherent state wave packet is constructed by
displacing the vacuum state $\vert 0 \rangle$ as
described in Equation~(\ref{disp}).
In case of non-linearity
proportional  to $(a^{\dag}a)^3$,  the dynamics of
coherent wave packet reveals the existence of
revivals as well as super-revivals of the
expectation values of oscillator amplitude
operator $\langle a\rangle$. In this case, the
revival time depends upon the principle quantum
number $n$. Therefore we also study the
effect of dissipative medium on the revivals of
coherent wave packet corresponding to displaced
number states for different values of $n$ as
defined in Equation~(\ref{dns}).  The analysis
of displaced number states is important only in
the case of higher-order non-linearity.

As a first computation, we consider the effect of
dissipation on the linear oscillator without any non-linear
terms. The results  are displayed in Fig.~\ref{fig-1} where
we have plotted  the amplitude  $\langle a \rangle$ as a
function of time. The simulation has been done for $59$
classical time periods which is equivalent to one revival
time $T_{\rm rev}$ corresponding to $b_1=0.008$ a.u,  where
we have taken the dissipation parameter $\gamma=0.01$ a.u.
In Fig.~\ref{fig-1} and in all subsequent figures we have
divided the real time by $T_{\rm rev}$ to have a uniform
scale.  The energy spectrum is equidistant and there is only
one time associated with motion namely the classical time
$T_{\rm cl}$.  Dissipation caused by environment is evident
through the damping and ultimate decay of the expectation
value of coherent wave packet amplitude $\langle a\rangle$
as time evolves.  The decay of amplitude is exponential as
expected.
\subsection{Results for non-linearity proportional
to $(a^{\dag}a)^2$}
\label{sec52}
For this case the results are presented in
Fig.~\ref{fig-2} where we display the expectation
value of the amplitude $\langle a\rangle$ as a
function of time for different values of
dissipation.  The value of the non-linearity
parameter $b_{1}$ is taken to be $0.0005$ a.u. It is
evident from Fig.~\ref{fig-2}(a) that in the
absence of coupling between system and environment
(marked by zero value of $\gamma$) full revival of
amplitudes takes place with a periodicity of
$k\,T_{\rm rev}/2$ with $k = 1, 2, 3 \cdots$. At
these revival times all eigenstates accumulate a
phase of $2 \pi k$.  This behavior has already
been studied theoretically~\cite{selfr}
as well as experimentally~\cite{14jm}. For finite but
small damping parameter $\gamma=0.00001$  a.u. as shown
in Fig.~\ref{fig-2}(b), due to damping a reduction
in revived amplitude is noticeable although the
revivals do take place.  As the value of $\gamma$
increases the revivals become weaker and weaker as
seen in (Fig.~\ref{fig-2}(c) \&(d)). For a certain
value of $\gamma$, approximately equal to $0.001$ a.u.,
the revivals disappear altogether.
Our results on
the one hand show robustness of the revival
phenomena in a dissipative environment and on the
other hand point towards a threshold level of
dissipation after which no revivals occur
whatsoever.

In Fig.~\ref{fig-3}, we study the effect of weak
and strong non-linearity on the revival behavior.
For zero non-linearity there are perfect revivals
with a time period of $T_{\rm cl}$.  Inclusion of
non-linear term in the Hamiltonian results in an
erratic energy spectrum leading to the collapse and
subsequent revival of
the wave packet. For very small values of the
non-linear parameter there is no perceptible effect
on the pattern as  seen in fig.~\ref{fig-3}(a)
where we have chosen a value $b_1=0.0000005$  a.u.
However, as seen from
fig.~\ref{fig-3}(b) the packet undergoes a revival
for small values of non-linearity  at time $T_{\rm
rev}$. As the value of the non-linearity parameter
is increased the revivals survive up to some value
(Fig.~\ref{fig-3}(c)) and then disappear
(Fig.~\ref{fig-3}(d)).  For Fig.~\ref{fig-3}(d),
the value of the non-linearity parameter is chosen
to be large ( $b_{1}=1.0$ a.u.) and the pattern is
irregular.
\begin{figure}
\includegraphics[scale=1]{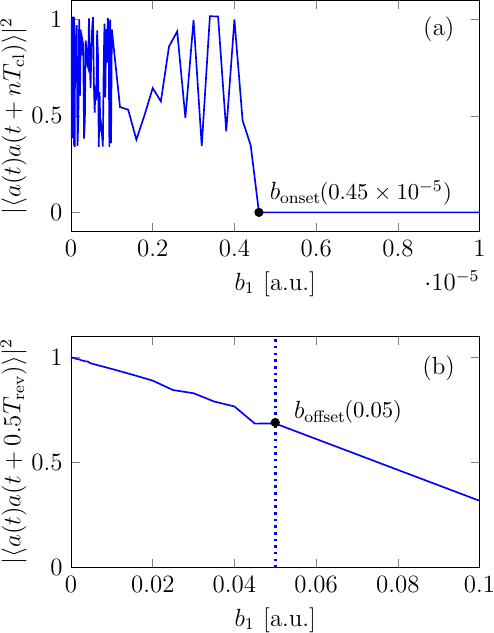}
\caption{Plots of correlation functions to ascertain 
the self-similarity of revival patterns. (a) The correlation
function is calculated with a time gap which is a multiple of
$T_{\rm cl}$ (we have taken $n=1500$). The sudden drop in this
correlation function indicates the onset of collapse at
$b_{\rm onset}$. (b) The correlation function between the
the revival patterns is plotted as a function of nonlinearity
parameter $b_1$ to ascertain the self-similarity of revival
pattern. It turns out that the correlation function decays
smoothly and we have to decide on a $b_{\rm offset}$ by visually
ascertaining the lack of similarity of the pattern.
\label{on_off}}
\end{figure}
As we scan different values of the non-linearity
parameter $b_1$,  we find that there is a threshold
value of $b_1$ for which the state actually
collapses and then revives at $T_{\rm rev}$. We
call this value of non-linearity parameter  as
$b_{\rm onset}$ and in our case $b_{\rm
onset}=0.0000045$ a.u. This value is ascertained from the
correlation function plot given in Fig.~\ref{on_off}(a).
Here it is very clear that the correlation function between
the values of the $\langle a \rangle$ separated by a time
which is a multiple of $T_{\rm cl}$ (We have taken $1500
T_{\rm cl}$) falls suddenly at this
value of the non-linear parameter $b_1$.  As the value of
the non-linearity parameter is further increased, the
revivals continue.  The correlation function between the
values  $\langle a \rangle$ separated by $0.5 T_{\rm rev}$
drops somewhat smoothly and linearly with the nonlinear
parameter $b_1$ as shown in Fig.~\ref{on_off}(b). Therefore,
there is no clear threshold after which we can say that the
revivals are not taking place. However, a visual inspection
of the graphs shows that they survive up to some value of
$b_1\approx 0.05$ (Fig.~\ref{fig-3}(c) and
Fig.~\ref{on_off}(b)) and then the pattern
looks different from the original (Fig.~\ref{fig-3}(d)).
We thus define non-linearity parameter above which the
revival pattern disappears as  $b_{\rm offset}$ and assign
it a value $0.05$ when the correlation function drops to
$0.7$.
\subsection{Presence of non decaying solutions}
\label{sec33}
\begin{figure}[h]
\includegraphics[scale=1]{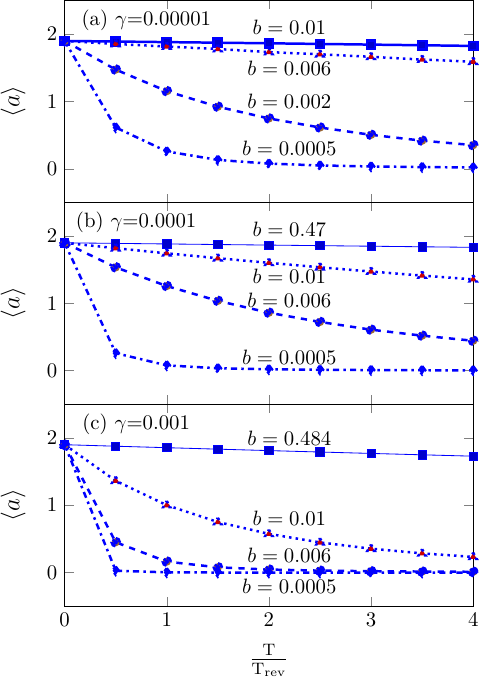}
\caption{
The plot $\langle a\rangle$ (real part) as a function of  time
period for coherent wave packet  passing through a
non-linear medium  for (a) $\gamma=0.00001$  a.u., (b)
$\gamma=0.0001$  a.u. and (c) $\gamma=0.001$  a.u. at various
values of non-linearity parameter. The uppermost
curve in all the cases pertains to
non-linearity compensating for dissipation leading
to persistent revival pattern.
}
\label{sol1}
\end{figure}

\begin{figure}[h]
\includegraphics[scale=1]{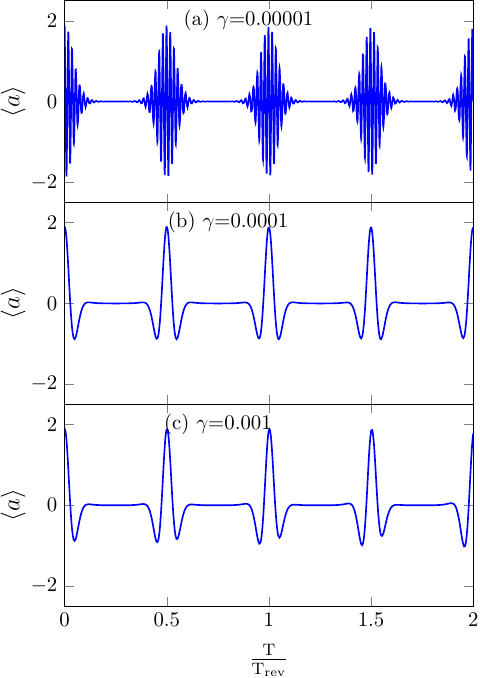}
\caption{
The plot of $\langle a\rangle$ (real part) as a function of time
on a fine grained time scale (time step much
smaller than $T_{rev}$) for a
coherent wave packet  passing through a non-linear
medium in the regime where dissipation is
compensated by non-linearity. Value of
non-linearity parameter in the soliton-like
behavior is $b_{S}= 0.01, 0.47, 0.484$ a.u.
in  (a), (b) and (c) respectively.
}
\label{sol2}
\end{figure}

Is there a possibility of having non-linearity
compensate dissipation leading to a 
solution where we have the excitation live for a
long time? It turns out that such a parameter
regime does exist as we have presented in
Fig.~\ref{sol1}. For a given value of the
dissipation we slowly increase the value of
non-linearity. We observe that for a given
value of dissipation constant we need to increase
the strength of non-linearity upto a certain
value to achieve a situation where revivals begin
to occur with full strength. This is very similar
to solitons of classical physics where we can have
non-linearity and dissipation acting in such a way
that a persistent waveform is generated. It is
interesting and surprising that such a situation
arises in the case of this fully quantum
treatment with dissipation modeled via the master
equation.
To observe the actual dynamics  in this case we choose the
non-linearity parameter in such a way that the soliton-like
solution is observed and then plot the expectation value of
$\langle a \rangle $ as a function of time on a time scale
much smaller than $T_{\rm rev}$.  We observe the revivals
with full strength upto several periods. This is depicted in
Fig.~\ref{sol2}. Furthermore, if nonlinearity is increased
further the compensation situation disappears again.
We would like to caution the reader that these solitons are
not of the variety where  dispersion is compensated by
nonlinearity and it is dissipation that is being compensated
and in the literature such situations have been called
solitons and therefore, following~\cite{ws} we call our
situation soliton like. Secondly, the presence of these
features is linked to the model of dissipation that we have
used and the actual physical relevance and explanation of
these results require further exploration.
\subsection{Results for non-linearity
proportional to $(a^{\dag}a)^3$}
\label{sec53}
\begin{figure}[h!]
\centering
\includegraphics[scale=1]{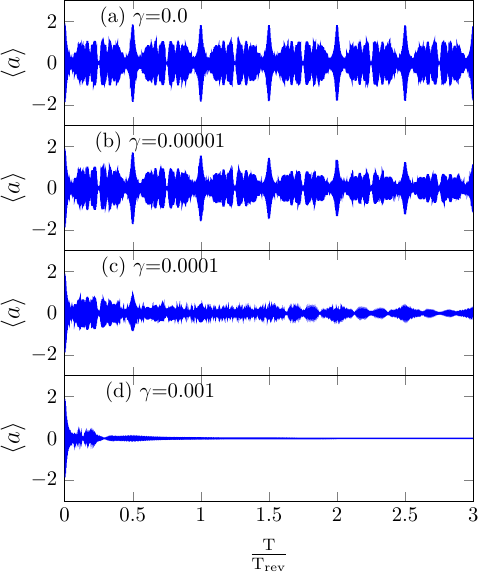}
\caption{Super-revival pattern of coherent state wave
packet.  Amplitude  $\langle a\rangle$ (real part) is plotted
as function of time in a medium
with non-linearity proportional to $(a^{\dag}a)^3$
for different values of dissipation values
$\gamma$. (a) $\gamma=0$, (b) $\gamma=0.00001$  a.u.,
(c) $\gamma=0.0001$   a.u. and  (d) $\gamma= 0.001$   a.u.}
\label{fig-4}
\end{figure}
\begin{figure}[h!]
\includegraphics[scale=1]{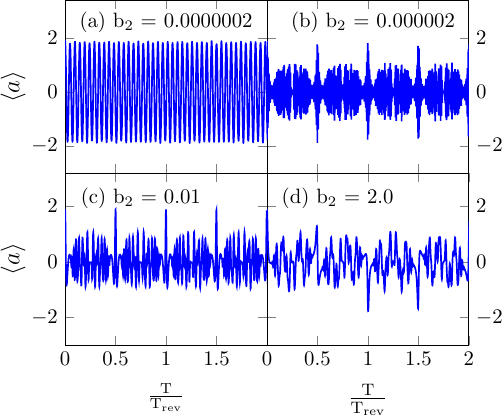}
\caption{Effect of weak and strong non-linearity
represented by parameter $b_2$, on the revival behavior
for  a coherent wave packet of red photon passing
through medium with non-linearity proportional to
$(a^{\dag}a)^3$ for various values of the
non-linearity parameter $b_2$.
(a) $b_2=0.0000002$   a.u.,  (b) $b_2=0.000002$  a.u.,
(c) $b_2=0.01$  a.u., and (d) $b_2=2.0$  a.u.}
\label{fig-5}
\end{figure}
To begin with we start with a coherent state
wave packet and study its behavior in the presence
of non-linearity and for different values of the
dissipation parameter as was done for the case of
Kerr non-linearity. Figures~\ref{fig-4} (a),(b),
(c) and (d) show the super-revival pattern of coherent
wave packet amplitude  $\langle a\rangle$ for
$\gamma=0,0.00001, 0.0001,$ and $0.001$ a.u., respectively.
The super-revival pattern is
evident in these figures.
Here the super-revival time $T_{\rm sr}$ is
independent of the principle quantum number $n$
(Equation~(\ref{tsr})) similar to the case studied in the
previous section where the revival time was
independent of $n$.
Within one super revival time some full revivals
and some partial revivals come into the picture.
The revival of the state occurs at the super
revival time which
is a longer time scale compared to the revival time.
For the sake of comparison with the
previous section we further study the effect of
strength of non-linearity in the system given by
the parameter value $b_2$.
From fig.~\ref{fig-5}, one can
conclude that for a medium with non-linear term
proportional to $(a^{\dag}a)^3$ revivals are
present up to a certain value of non-linearity
beyond which the pattern becomes irregular.  For
values of $b_2$ (figs.~\ref{fig-5}(b) \& (c)) the
revivals are present.  For $b_2$ values greater
then $0.01$ the revivals are irregular. For strong
non-linearity $b_2=2.0$ a.u., the revivals are not only
irregular but also almost stop appearing, which is
depicted in Fig.~\ref{fig-5}(d). As we discussed
for the earlier case we have the notions of
$b_{\rm onset}$ and $b_{\rm offset}$ here too.
Figure~\ref{fig-5}(a) $(b_2=0.0000002)$  a.u. pertains to
a value of non-linearity below $b_{\rm onset}$. In
this case too, from a scan through different values of $b_2$
we find that  $b_{\rm onset}=0.000002$ a.u. and $b_{\rm
offset}=0.01$ a.u. approximately.
\subsection{The effect of $n$}
The fact that revival time in this case shows
dependence on principle quantum number $n$
(Equation~(\ref{trev})), provides an opportunity
to explore the revival pattern for various values
$n$ for displaced number states.  As defined
in Equation~(\ref{dns})  we considered displaced
number states $\vert \alpha,n \rangle$.
\begin{figure}
\includegraphics[scale=1]{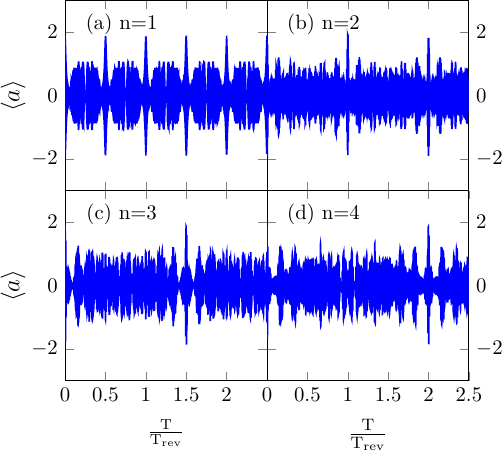}
\caption{Collapse and revival pattern of displaced
number states with amplitude  $\langle a\rangle$ (real part)
being plotted as a function of time with
non-linearity proportional to $(a^{\dag}a)^3$ and
no dissipation. The pattern is shown for states
constructed around $n=1,2,3,4$ at $\gamma=0$  a.u. The x-axis is
normalized with respect to theoretical $T_{\rm rev}$ which is  around
$4000$, $2000$, $1330$ and $1000$  a.u. for $n=1,2,3$ and
$4$  respectively.} \label{fig-6}
\end{figure}
\begin{figure}
\includegraphics[scale=1]{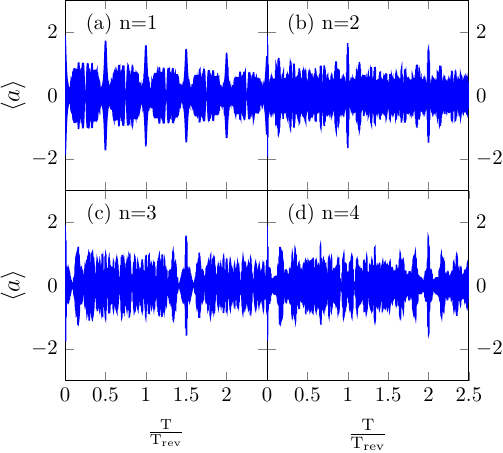}
\caption{Collapse and revival pattern of displaced
number states with amplitude  $\langle a\rangle$ (real part)
being plotted as a function of time with
non-linearity proportional to $(a^{\dag}a)^3$ and
in the presence of  dissipation. The pattern is shown for states
constructed around $n=1,2,3,4$ at $\gamma=0.00001$  a.u.  The x-axis is
normalized with respect to theoretical $T_{\rm rev}$ which is  around
$4000$, $2000$, $1330$ and  $1000$ a.u. for $n=1,2,3$
and $4$  respectively.}
\label{fig-7}
\end{figure}

The results of this study are shown in
Figs.~\ref{fig-6} and~\ref{fig-7}.  It is seen
that the revival  patterns for displaced number
states are dependent upon $n$.  We have taken the
values $n=1,2,3,4$, with non-linearity
$b_{2}=0.0005$ a.u., in the absence and presence of the
system-environment coupling, respectively.
Fig.~\ref{fig-6} represents the results without
dissipation and Fig.~\ref{fig-7} represents the
results in the presence of dissipation.  The
revival pattern is clearly affected by the
dissipative effects.

Theoretically calculated revival times for
$n=1,2,3$ and $4$ are around $4000, 2000, 1330$ and
$1000$ a.u. respectively.  However, from
Fig.~\ref{fig-6}(a), one can notice that the
revival of the wave packet occurs at half of
the theoretical revival time. This behavior for
the $n=1$
case was discussed in~\cite{selfr}
and~\cite{14jm}, where it was shown that revivals
appear only when all eigenstates have accumulated
a phase of $2 \pi k$ and in the process some extra
revivals may appear.  However, from our
Figs.~\ref{fig-6}(c)\& (d), we also observe that
there are cases where some revivals are missing
for higher $n$ values. It is evident from these
sub-figures that revivals of wave packets occur at
$2000$ a.u. independent of the $n$ value
for the state $\vert \alpha, n \rangle$.

 A comparison of
Fig.~\ref{fig-6} with Fig.~\ref{fig-7} also
indicates that the wave packet constructed around
the higher displaced states are dissipated more
prominently.  In order to elucidate this finding,
we extend the value of $n$ for the state $\vert
\alpha, n\rangle $ upto $n=10$ and study the
effect of dissipative medium at the first revival
time.

\begin{figure}[h!] \centering
\includegraphics[scale=1]{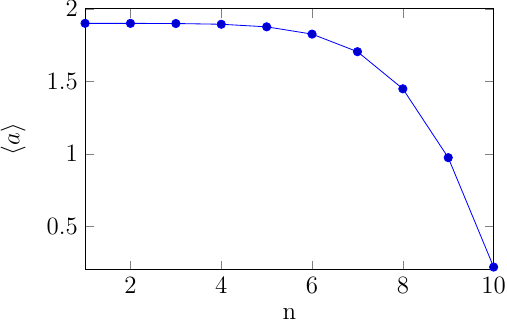}
\caption{Effect of dissipative medium on the
displaced number states  constructed around
various $n$  in a medium with non-linearity
proportional to $(a^{\dag}a)^3$. The amplitude of
revived wave packets at their first revival time
instant are recorded for various values of $n$.
The dissipation of amplitude is more vigorous for
higher values of $n$.}
\label{fig-8}
\end{figure}

The results are shown  in Fig.~\ref{fig-8}, where the
displaced number states constructed for higher $n$
witness more dissipation. This is an 
interesting observation which is related to the fact
that the displaced number states with higher $n$
are more nonclassical and hence the effect of
dissipation is rapid.  Similar results were found
for a related phenomenon of decoherence
in~\cite{pfoldi}.  `
\section{Concluding Remarks}
\label{concl}
In this work the quantum master equation approach
has been used to analyze the dissipative dynamics
of a coherent wave packet passing through
non-linear media of two different types.  The
effect of a dissipative bath on the phenomena of
collapses and revivals is evident. However, it is
observed that the basic features of the revival
process survives some amount of dissipation.
Although the revival is never complete in the
presence of dissipation, its signatures are
present. This is relevant for any attempt to
observe these revivals experimentally.

The other prominent role played  by the
dissipative medium is that of an environment which
soaks the energy of the wave packet.  Dissipation
is observed in the damping of successive revived
coherent wave packet amplitudes  $\langle
a\rangle$. The controlling element for dissipation
is the damping factor $\gamma$ which further
depends upon the coupling between the coherent
wave packet and the medium.  With increase in the
strength of damping factor $\gamma$, the amplitude
of the successively revived wave packet decreases
and ultimately no revivals occur for a particular
value of $\gamma$.

A particularly interesting situation arises for a
certain range of parameters where the non-linearity
and dissipation are poised in such a way that the
revivals occur even in the presence of
dissipation for a long time. In this regime of parameters, the
amplitude of the revivals remains constant and
there is no loss of energy. This is very similar to
the phenomena of solitons in classical non-linear
waves in the presence of dissipation. We believe
this observation may have useful consequences for
quantum communication protocols where one is
interested in fabricating low dissipation
communication channels.

For a medium with non-linearity proportional to
$({a}^{\dag} a)^{3}$, signatures of super revivals
are evident. Theoretically, revivals are predicted
at different times for wave packets constructed
around different displaced quantum numbers $n$.
However, we observe that the revival of the wave
packet occurs at a fixed time independent of the
$n$. This behavior sheds light on a very
fundamental aspect that revivals appear only when
all eigenstates have accumulated a phase of $2 \pi
k$ and in the process some extra revivals may
appear and some predicted ones may be missing.

We also observe that the wave packet build around the higher
values of principle quantum number are more prone to the
dissipative effects of the medium as compared to the low
lying states. The extent of non-linearity of the medium also
has an effect on the dynamics of the wave packet.  We come
across a value of the non-linearity parameter $b_{\rm
onset}$ below which no collapses and hence no revivals
occur. Similarly, there is an  upper limit of $b$ that we
call  $b_{\rm offset}$, above which the pattern of revivals
is irregular.

We have demonstrated that revivals can tolerate certain
amount of dissipation, this certainly is useful in observing
this phenomena in the lab. The soliton like solutions that
we have found if experimentally realized may turn out to be
useful in many practical situations. However, more work is
needed to evaluate the relevance and possible applications
of these results.
\begin{acknowledgments}
B.A.  acknowledges  DST, INDIA for funding under
Grant No. EMR/2016/001228. Arvind
acknowledges DST India for funding under Grant No.
EMR/2014/000297.
\end{acknowledgments}
%

\end{document}